\documentstyle[twoside,epsfig]{article}

\catcode`\@=11
\long\def\@makefntext#1{
\protect\noindent \hbox to 3.2pt {\hskip-.9pt
$^{{\eightrm\@thefnmark}}$\hfil}#1\hfill}               

\def\@makefnmark{\hbox to 0pt{$^{\@thefnmark}$\hss}}    

\def\ps@myheadings{\let\@mkboth\@gobbletwo
\def\@oddhead{\hbox{}
\rightmark\hfil\eightrm\thepage}
\def\@oddfoot{}\def\@evenhead{\eightrm\thepage\hfil
\leftmark\hbox{}}\def\@evenfoot{}
\def\sectionmark##1{}\def\subsectionmark##1{}}

\oddsidemargin=\evensidemargin
\newcounter{sectionc}\newcounter{subsectionc}\newcounter{subsubsectionc}
\renewcommand{\section}[1] {\vspace{12pt}\addtocounter{sectionc}{1}
\setcounter{subsectionc}{0}\setcounter{subsubsectionc}{0}\noindent
        {\tenbf\thesectionc. #1}\par\vspace{5pt}}
\renewcommand{\subsection}[1] {\vspace{12pt}\addtocounter{subsectionc}{1}
        \setcounter{subsubsectionc}{0}\noindent
        {\bf\thesectionc.\thesubsectionc. {\kern1pt \bfit #1}}\par\vspace{5pt}}
\renewcommand{\subsubsection}[1] {\vspace{12pt}\addtocounter{subsubsectionc}{1}
        \noindent{\tenrm\thesectionc.\thesubsectionc.\thesubsubsectionc.
        {\kern1pt \tenit #1}}\par\vspace{5pt}}
\newcommand{\nonumsection}[1] {\vspace{12pt}\noindent{\tenbf #1}
        \par\vspace{5pt}}

\newcounter{appendixc}
\newcounter{subappendixc}[appendixc]
\newcounter{subsubappendixc}[subappendixc]
\renewcommand{\thesubappendixc}{\Alph{appendixc}.\arabic{subappendixc}}
\renewcommand{\thesubsubappendixc}
        {\Alph{appendixc}.\arabic{subappendixc}.\arabic{subsubappendixc}}

\renewcommand{\appendix}[1] {\vspace{12pt}
        \refstepcounter{appendixc}
        \setcounter{figure}{0}
        \setcounter{table}{0}
        \setcounter{lemma}{0}
        \setcounter{theorem}{0}
        \setcounter{corollary}{0}
        \setcounter{definition}{0}
        \setcounter{equation}{0}
        \renewcommand{\thefigure}{\Alph{appendixc}.\arabic{figure}}
        \renewcommand{\thetable}{\Alph{appendixc}.\arabic{table}}
        \renewcommand{\theappendixc}{\Alph{appendixc}}
        \renewcommand{\thelemma}{\Alph{appendixc}.\arabic{lemma}}
        \renewcommand{\thetheorem}{\Alph{appendixc}.\arabic{theorem}}
        \renewcommand{\thedefinition}{\Alph{appendixc}.\arabic{definition}}
        \renewcommand{\thecorollary}{\Alph{appendixc}.\arabic{corollary}}
        \renewcommand{\theequation}{\Alph{appendixc}.\arabic{equation}}
        \noindent{\tenbf Appendix \theappendixc #1}\par\vspace{5pt}}
\newcommand{\subappendix}[1] {\vspace{12pt}
        \refstepcounter{subappendixc}
        \noindent{\bf Appendix \thesubappendixc. {\kern1pt \bfit #1}}
        \par\vspace{5pt}}
\newcommand{\subsubappendix}[1] {\vspace{12pt}
        \refstepcounter{subsubappendixc}
        \noindent{\rm Appendix \thesubsubappendixc. {\kern1pt \tenit #1}}
        \par\vspace{5pt}}

\newcommand{\textlineskip}{\baselineskip=13pt}
\newcommand{\smalllineskip}{\baselineskip=10pt}

\def\eightcirc{
\begin{picture}(0,0)
\put(4.4,1.8){\circle{6.5}}
\end{picture}}
\def\eightcopyright{\eightcirc\kern2.7pt\hbox{\eightrm c}}

\newcommand{\copyrightheading}[1]
        {\vspace*{-2.5cm}\smalllineskip{\flushleft
        {\footnotesize $\eightcopyright$\, World Scientific Publishing
         Company}\\
         }}

\def\abstracts#1#2#3{{
        \centering{\begin{minipage}{4.5in}\baselineskip=10pt\footnotesize
        \parindent=0pt #1\par
        \parindent=15pt #2\par
        \parindent=15pt #3
        \end{minipage}}\par}}

\newcommand{\bibit}{\nineit}
\newcommand{\bibbf}{\ninebf}
\renewenvironment{thebibliography}[1]
        {\frenchspacing
         \ninerm\baselineskip=11pt
         \begin{list}{\arabic{enumi}.}
        {\usecounter{enumi}\setlength{\parsep}{0pt}
         \setlength{\leftmargin 12.7pt}{\rightmargin 0pt} 
         \setlength{\itemsep}{0pt} \settowidth
        {\labelwidth}{#1.}\sloppy}}{\end{list}}

\newcounter{itemlistc}
\newcounter{romanlistc}
\newcounter{alphlistc}
\newcounter{arabiclistc}

\newcommand{\fcaption}[1]{
        \refstepcounter{figure}
        \setbox\@tempboxa = \hbox{\footnotesize Fig.~\thefigure. #1}
        \ifdim \wd\@tempboxa > 5in
           {\begin{center}
        \parbox{5in}{\footnotesize\smalllineskip Fig.~\thefigure. #1}
            \end{center}}
        \else
             {\begin{center}
             {\footnotesize Fig.~\thefigure. #1}
              \end{center}}
        \fi}

\newcommand{\tcaption}[1]{
        \refstepcounter{table}
        \setbox\@tempboxa = \hbox{\footnotesize Table~\thetable. #1}
        \ifdim \wd\@tempboxa > 5in
           {\begin{center}
        \parbox{5in}{\footnotesize\smalllineskip Table~\thetable. #1}
            \end{center}}
        \else
             {\begin{center}
             {\footnotesize Table~\thetable. #1}
              \end{center}}
        \fi}

\def\@citex[#1]#2{\if@filesw\immediate\write\@auxout
        {\string\citation{#2}}\fi
\def\@citea{}\@cite{\@for\@citeb:=#2\do
        {\@citea\def\@citea{,}\@ifundefined
        {b@\@citeb}{{\bf ?}\@warning
        {Citation `\@citeb' on page \thepage \space undefined}}
        {\csname b@\@citeb\endcsname}}}{#1}}

\newif\if@cghi
\def\cite{\@cghitrue\@ifnextchar [{\@tempswatrue
        \@citex}{\@tempswafalse\@citex[]}}
\def\citelow{\@cghifalse\@ifnextchar [{\@tempswatrue
        \@citex}{\@tempswafalse\@citex[]}}
\def\@cite#1#2{{$\null^{#1}$\if@tempswa\typeout
        {IJCGA warning: optional citation argument
        ignored: `#2'} \fi}}

\def\pmb#1{\setbox0=\hbox{#1}
        \kern-.025em\copy0\kern-\wd0
        \kern.05em\copy0\kern-\wd0
        \kern-.025em\raise.0433em\box0}

\def\fnm#1{$^{\mbox{\scriptsize #1}}$}
\def\fnt#1#2{\footnotetext{\kern-.3em
        {$^{\mbox{\scriptsize #1}}$}{#2}}}

\def\fpage#1{\begingroup
\voffset=.3in
\thispagestyle{empty}\begin{table}[b]\centerline{\footnotesize #1}
        \end{table}\endgroup}

\def\runninghead#1#2{\pagestyle{myheadings}
\markboth{{\protect\footnotesize\it{\quad #1}}\hfill}
{\hfill{\protect\footnotesize\it{#2\quad}}}}
\font\tenrm=cmr10
\font\tenit=cmti10
\font\tenbf=cmbx10
\font\bfit=cmbxti10 at 10pt
\font\ninerm=cmr9
\font\nineit=cmti9
\font\ninebf=cmbx9
\font\eightrm=cmr8

\def\qed{\hbox{${\vcenter{\vbox{                        
   \hrule height 0.4pt\hbox{\vrule width 0.4pt height 6pt
   \kern5pt\vrule width 0.4pt}\hrule height 0.4pt}}}$}}

\def\bsc{{\sc a\kern-6.4pt\sc a\kern-6.4pt\sc a}}       
\def\bflatex{\bf L\kern-.30em\raise.3ex\hbox{\bsc}\kern-.14em
T\kern-.1667em\lower.7ex\hbox{E}\kern-.125em X}

\begin{document}

\normalsize\textlineskip
\thispagestyle{empty}
\setcounter{page}{1}



{\large
\begin{flushright}LAPP-EXP-95.06\end{flushright}
\vspace*{0.88truein}
\begin{center}{\bf  {\Large Towards a Complete Feynman Diagrams Automatic
Computation System}}
\end{center}
\vspace*{0.035truein}
\vspace*{0.37truein}
\centerline{ D. PERRET-GALLIX}
\vspace*{0.1truein}
\centerline{\it L.A.P.P., IN$^2$P$^3$, CNRS}
\centerline{\sc e-mail: \it perretg@cernvm.cern.ch}
\baselineskip=10pt
\centerline{\it B.P. 110 74941 Annecy-Le-Vieux  France,
}
\vspace*{10pt}

\vspace*{0.4truein}
\begin{center} \bf Abstract: \end{center}
\begin{quote}
Complete Feynman diagram automatic computation systems are
now coming of age after many years of development. They are
made available to the high energy physics community through user-friendly
interfaces.
Theorists and experimentalists can benefit
from these powerful packages for speeding up time consuming calculations
and for preparing event generators.
The general architecture of these packages is presented and
the current development of the one-loop diagrams extension is
discussed.
A rapid description of the prominent packages and tools is
then proposed. Finally, the necessity for defining a standardization
scheme is heavily
stressed for the benefit of developers and users.
\end{quote}



\vspace*{1pt}\textlineskip      
\vspace*{3cm}
\begin{center}
\it Invited talk at the Fourth International Workshop on Software Engineering
and
Artificial Intelligence for High Energy and Nuclear Physics. \\
AIHENP-95 \\
Pisa April 3-8 1995
\end{center}
}
\eject
{}~~
\eject
\runninghead{Towards a Complete Feynman Diagrams Automatic Computation
System} {Towards a Complete Feynman Diagrams Automatic Computation
System}

\fpage{1}
\centerline{\bf Towards a Complete Feynman Diagrams Automatic
Computation System}
\vspace*{0.035truein}
\vspace*{0.37truein}
\centerline{\footnotesize D. PERRET-GALLIX
}
\vspace*{0.015truein}
\centerline{\footnotesize\it L.A.P.P., IN$^2$P$^3$, CNRS}
\centerline{\footnotesize\sc e-mail: \it perretg@cernvm.cern.ch}
\baselineskip=10pt
\centerline{\footnotesize\it B.P. 110 74941 Annecy-Le-Vieux  France,
}
\vspace*{10pt}

\vspace*{0.21truein}
\abstracts{
Complete Feynman diagram automatic computation systems are
now coming of age after many years of development. They are
made available to the high energy physics community through user-friendly
interfaces.
Theorists and experimentalists can benefit
from these powerful packages for speeding up time consuming calculations
and for preparing event generators.
The general architecture of these packages is presented and
the current development of the one-loop diagrams extension is
discussed.
A rapid description of the prominent packages and tools is
then proposed. Finally, the necessity for defining a standardization
scheme is heavily
stressed for the benefit of developers and users.
}{}{}



\vspace*{1pt}\textlineskip      
\section{Introduction}    
\vspace*{-0.5pt}
\noindent
Since its introduction by R.P Feynman\cite{feynman} in 1949, the diagrammatic
technique of computing matrix element and consequently, all physics quantities
in
high
energy physics (HEP) has been extensively used and has proved to be the most
simple,
intuitive and general method to compute even the most complex processes.
For its conciseness and its pictorial approach, this method has spread over
other
research field like atomic, nuclear and solid state physics.

\begin{figure}
\vfill \begin{minipage}{.495\linewidth}
\flushleft{\fbox{\epsfig{file=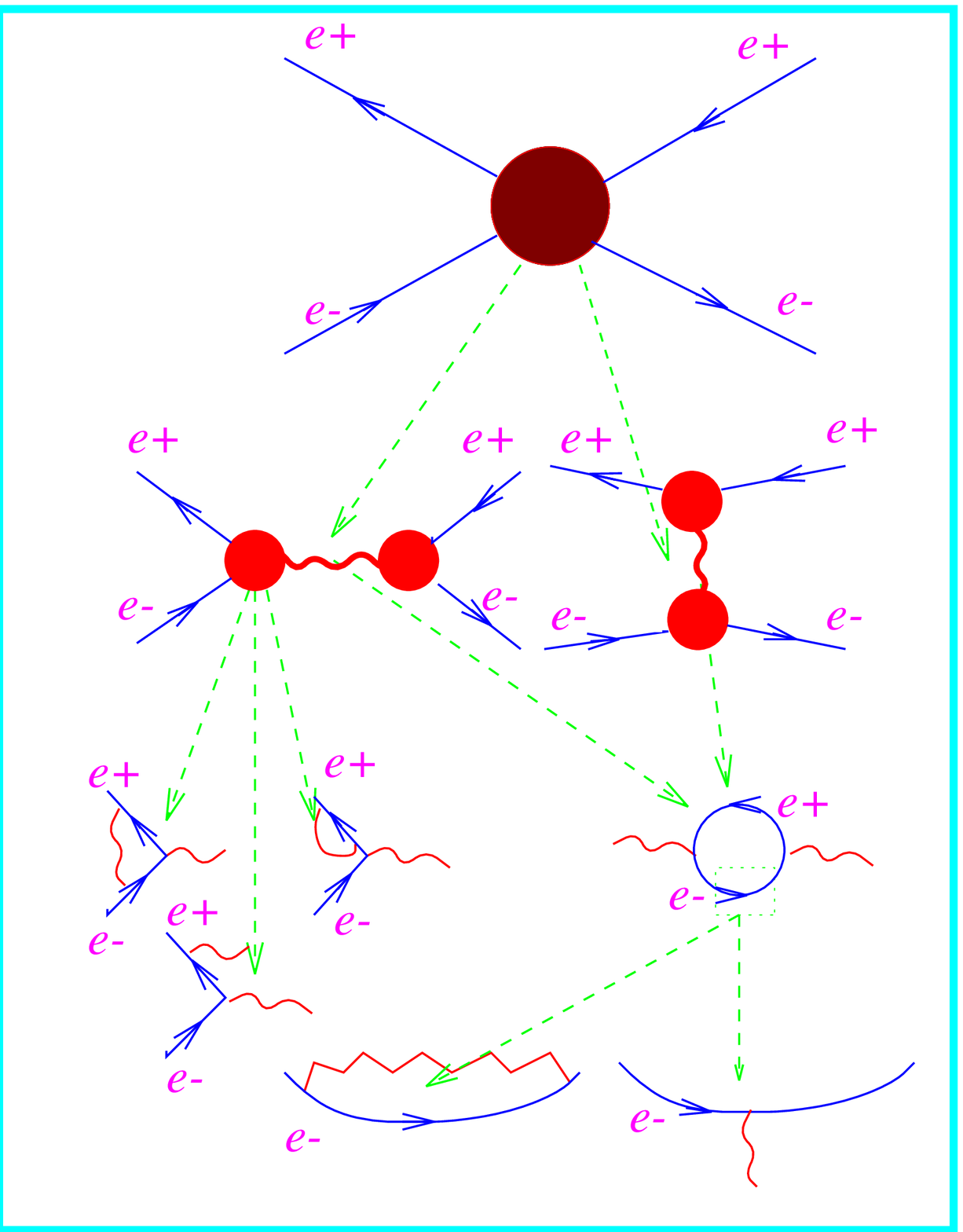,
height=7.5cm}}
\vspace*{0.7cm}
\fcaption{Decomposition of a process in Feynman diagrams} \label{fig:feynman}
}
\end{minipage} \hfill
 \begin{minipage}{.495\linewidth}
\flushright{\fbox{\epsfig{file=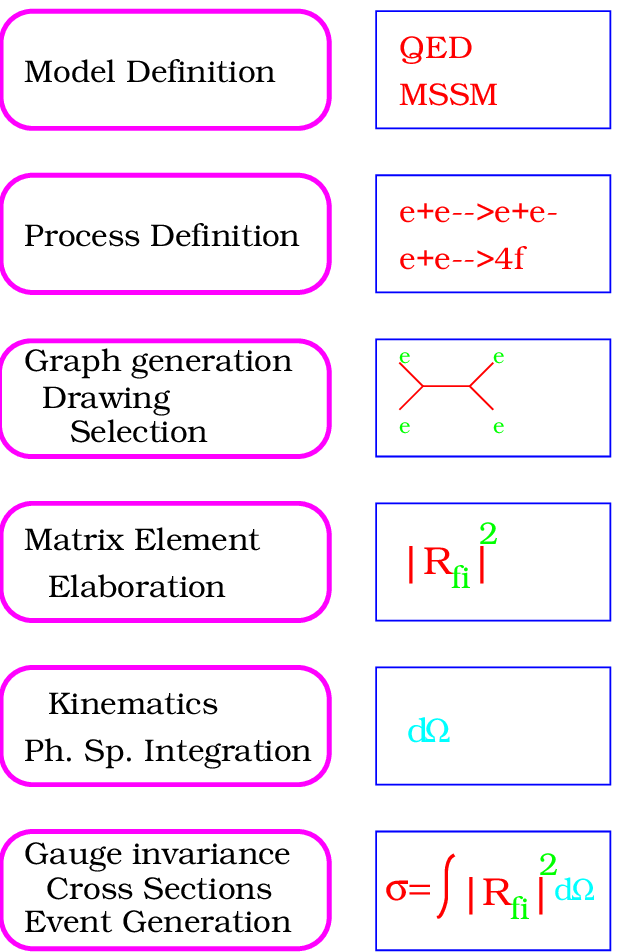,
height=7.5cm}}
\vspace*{0.3cm}
\fcaption{Main modules of a typical package} \label{fig:struct}
}
\end{minipage}
\vfill
\vspace*{-0.5cm}
\end{figure}

Once the Lagrangian of the theory is selected,
a process (defined by the initial and final state particles and
the order of the calculation) is decomposed into a set of sub-processes
represented by a diagram (fig.\ref{fig:feynman}).
The real power of the Feynman approach lies, through the use of definite
rules, on a quasi-mechanical transformation of each diagram in an algebraic
expression representing its quantitative contribution to the process.
Finally, the total cross-section is obtained by the
integration over the phase space of
the square of the sum of the contribution of all individual graphs.
Although this approach is straight forward in its principles, the actual
computation
of all but the most simple processes is a quite lengthy task, prone to
errors and mistakes. In order to avoid these pitfalls, calculus were used to be
performed independently by
several theorists until the results did fully agree.
In 1967 M. Veltman\cite{veltman} wrote the first computer language
to perform algebraic computation of the trace expressions
resulting from the Feynman diagrams method.
Schoonschip\cite{schoonschip} was the first step
towards an automatization of these quite involved computations.
For solving this same HEP computation problems, several other developments
started
at this time: {\it Reduce\cite{reduce0}, Macsyma\cite{macsyma0}}.
Nowadays general purpose computer algebra languages have been put on the
marketplace. They provide valuable tools to perform many physics
calculations\fnm{\dag}\fnt{\dag}{P.
Nason's paper in these proceedings}:
{\it Reduce\cite{reduce}, Maple\cite{maple}, Mathematica\cite{mathematica}}.
However, the original
{\it Schoonschip} and its more modern implementation, the {\it Form}
language\cite{form}
(J. Vermaseren) are still prefered for these specialized computations.
The technique developed by R.P. Feynman is well structured and quite suited
to a complete automatization. However the
idea of building a complete package performing the computation of any given
process only from general principles
appeared to many as an enormous task and a never ending enterprise. But several
groups out-passing these arguments decided to launch such projects.
This workshop series has played a positive role in the motivation and
the building of collaboration between these groups. Today,
several packages can be used by experimentalists to create (yet with
some limitations) their own event generator in an almost automatic way. Tools
and
libraries have also been created to facilitate complex calculations.
\section{Motivations}
\noindent
These packages which could have remained at the level of pedagogical toys,
have attracted a lot of interest because modern HEP research requires
the calculation of more and more complex processes.

\subsection{Larger center of mass energy}
\noindent
The center of mass energy of the colliders going up (LEP-II, HERA,
LHC and NLC), new heavy particle thresholds (W, Z , top, Higgs) are crossed
increasing therefore the number of diagrams and the level of complexity of each
diagram computation.

\subsection{Better accuracy}
\noindent
High precision experiments require high precision theoretical predictions.
$e^+e^-$ colliders experiments belong to this category and
channels bearing small contributions must be included in the computation to
cope
 with the
high statistic data.
For example, the
t-channel in the $e^+e^- \to l l \gamma \gamma$\cite{llgg} where intringing
event
accumulation for large M$_{\gamma\gamma}$ was observed in L3. The contribution
of
the non-resonant
t-channels in the W production at LEP-II\cite{singlew}, often neglected, must
be
computed below and well above the W-pair threshold.
Higher precision also means including loops diagrams as for the
luminosity measurement at LEP-I which is limited by the theoretical
uncertainty of the 2-loop calculation of the Bhabha scattering
($\Delta_{th}\approx 0.25\%,\Delta_{exp}\approx 0.07-0.16\%$).
The g-2 experiments
requires the computation of 4-loops contributions\cite{kinoshita}
to barely match
the experimental
precision.
Loop calculations are more involved than tree level ones
as the n-dimensional regularization technique must be applied
to treat the singularities.
Furthermore,
better accuracy imposes often to take into account fermion mass effects,
radiative
corrections and polarization effects.

\subsection{More Processes}
\noindent
Looking for small effects revealing possible deviations from the standard model
leads to the precise computation of many more processes: the "background
processes",
of no great theoretical importance
but whose contribution could dilute or completely wash out the expected signal.
Furthermore in the so-called supersymmetric models, the final state particles
are often unstable ($\tilde{q} \to q + \tilde{\gamma}$,
$\tilde{\chi}^+ \to \tilde{\chi}^o +W^+ \to \tilde{\chi}^o q \bar{q}$) and
therefore the actual processes are rather complex (2$\to$ 4,6 or more
particles).
Due to the particle inflation, more propagators are present.
For example, $e^+e^- \to e^+ e^- \tilde{\chi}^o \tilde{\chi}^o$ includes 704
diagrams
not taking into account the $\tilde{\chi}^o$ decay.
Moreover, there are many possible models
and each process may need to be computed for each model.

\subsection{Perturbative QCD}
\noindent
Next to leading order QCD computations are highly needed for a precise
comparison with the experiments, but the
complexity of loop calculation in non-abelian gauge theories
bridles this expectation. New methods beyond the conventional Feynman
diagrammatic techniques are being developed including spinor helicity methods,
supersymmetry Ward identities or string-based techniques\cite{bern}. The
inclusion of these
new approaches in automatic computation packages is the next logical step.
%

In summary, each diagram computation gets more complex (polarization, mass
effects, loops),
each process gets more diagrams (loops, non-standard models, non-resonant
channels), for each
model, more processes (background processes,..) must be computed and finally
more models have to be investigated. Therefore the need for automatic
Feynman computation packages is clear as this task is beyond
the reach of the theorist community.

\section{Components of an Feynman diagrams automatic computation package}
\noindent
 The general structure of a typical package is represented in the
fig.\ref{fig:struct}.

\subsection{Model definition}
\noindent
The framework on which the computation takes place has to be precisely and
uniquely defined. In principle the selection of the Lagrangian should lead
to the specification of all coupling constants. All fundamental parameters are
computed
from a set of
experimental values. Several models have already been implemented, including
QED, QFD, QCD and
MSSM\fnm{\dag}\fnt{\dag}{Quantum Electro Dynamic, Quantum Electroweak Dynamic,
Quantum Chromo Dynamic and
Minimal Supersymmetric Standard Model}.
For example, QFD can be defined\fnm{\ddag}\fnt{\ddag}{LEP-II
workshop proposal (see R. Kleiss's talk)}
from a selection of the following parameters
($\alpha(0), \alpha(Q^2), \alpha_s(Q^2),$ $ M_Z, \Gamma_Z,$
$ M_W, \Gamma_W, G_F,$ $\sin^2\theta_w,
V^{ij}_{CKM}, M_H, M_f$) depending essentially on the
experimental precision. The coupling parameters can be computed
($Q_f, g^f_v, $ $g^f_a, g^f_W, g_{WWZ}$) as well as $M_W$ and $\Gamma_W$
the W propagator mass and width.
Then the final observables
($|R|^2,\frac{d\sigma}{d\cos\theta}, \sigma_{tot}$)
should
be uniquely defined.

\subsection{Process definition}
\noindent
The initial and final state particles are selected as well as the
order in the coupling constant ($\alpha, g, \alpha_s$). For example:
$e^+e^-\to W^+W^-\gamma$ at $\cal{O}$$(\alpha^3)$ for tree level or
$e^+e^-\to W^+W^-$ at $\cal{O}$$(\alpha^4)$ for 1 loop corrections. The final
state
can be generic like $e^+e^-\to 4$ fermions or
$\gamma \gamma \to q\bar{q}$.
The initial state may contain composite particles as in
$P \gamma \to P \gamma$. The way to handle
radiative correction (ISR, soft photon contribution) and structure functions
should also be defined.

\subsection{Graph generation, drawing and selection}
\noindent
At the tree level, the graph generation is straight forward, the generalization
to the n-loop case has been solved by the so-called "orderly
algorithm"\cite{orderly}
where all possible topologies without duplication are identified.
However the computing time at a given order $\cal{O}$$(\alpha^n)$ is
proportional to the \#nodes!\fnm{\S}\fnt{\S}{A node is an external particle
or a vertex}.
Performance improvement for complex event has been achieved\cite{kaneko} based
on the use of vertex classification.

The graph drawing module, producing usually postscript diagram representation,
is needed for
visual check, interactive selection and inclusion in
publication.
This work of art has been completed in some case up to
2-loop diagrams where topology difficulties becomes
quite substantial.

Diagram ordering or classification based on some
physics interest
is a very important issue.
Finding rules based on topological informations and model properties to select
gauge invariant graph subgroups or to order diagrams by their expected
contributions would provide ways to reduce computation loads.

\subsection{Matrix element elaboration}
\noindent
Having generated all the diagrams and using the model
description, one can now write the matrix element function. It
will provide the contribution of all selected diagrams for each phase space
point. Two approaches are possible and have been used.
Let us consider, for example, the inclusive reaction, $e^+e^- \to X$.
The matrix element can be written as follows:
\begin{equation}
|R|^2=\sum_{\scriptstyle\xi,\xi'}|\sum_{\scriptstyle
g}\bar{v}(\ell',\xi')\vartheta
u(\ell,\xi)|^2
\label{equ:1}
\end{equation}
where $\xi,\xi'$ are helicities and $g$ graphs. \\
{\bf The symbolic approach}
\begin{equation}
|R|^2=\sum_{\scriptstyle
g,g'}\sum_{\scriptstyle\xi,\xi'}(\bar{v}(\ell',\xi')\vartheta
u(\ell,\xi))(\bar{v}(\ell',\xi')\vartheta
u(\ell,\xi))^*
\end{equation}
\begin{equation}
|R|^2=\sum_{\scriptstyle g,g'}Tr(\rho'\vartheta_g\rho\tilde{\vartheta_{g'}})
 ~~{where} ~~\tilde{\vartheta}=\gamma^0\vartheta^\dag\gamma^0
\end{equation}
is based on trace calculus using symbolic manipulation languages: {\it
Schoonschip, Reduce, Form, Maple, Mathematica.}
The sum of the trace for all graphs is performed after summing over
the helicities of the external particles.
This technique leads to
compact expressions for simple processes and to the analytic cancelation of
singularities.
However large expressions are difficult to reduce (nb. terms $\propto$ (nb.
of diagrams)$^2$), polarization effect cannot easily be computed as it adds
many more terms
and the gauge invariant check is difficult as the
computation can only been done in the most appropriate gauge scheme.
\\
{\bf In the numerical approach},
expression (\ref{equ:1}) is directly developed in wave functions,
vertices and propagators components.
The matrix element is then built from sequential calls to a helicity
amplitude library containing all necessary numerical routines associated to
each basic components.
This is a systematic approach, valid for all tree level processes and not
limited by
the complexity of the diagrams.
However, numerical stability hampers the practical use of this technique.
The load is put on the
integration package and its ability to deal with singularities.
Large computing time is needed for the integration and event generation. For
example
100 hours on an HP-735 were necessary to compute a
(2 $\to$ 5) process like $e^+e^-\to e^-\bar{\nu_e}u\bar{d}\gamma $.

\subsubsection{Higher order computation}
\noindent
Three cases should be considered: 1-loop, 2-loop and n-loop corrections.
In the first case, a mixed approach (symbolic and numerical) is followed.
Tree level, 1-loop diagrams and counterterms
are generated, then the product of loop and tree diagrams is performed
symbolically: $\Sigma_{_{i,j}}(loop)_i*(tree)^*_j$. The output is composed of
2-4 point functions. A numerical integration is then performed using
numerical\cite{grace-t} or analytical\cite{olden} scalar loop integral
libraries( up-to 5-point functions).
Although limited to 4 external particles in QFD, automatic
computation have been used for $e^+e^- \to t \bar{t}$ (2 tree and 50 1-loop
diagrams) or $e^+e^- \to Z^0 H$ (1 tree and 80 1-loop diagrams).
Numerical instabilities and computer performance are the major limitations
of this approach.
Two-loop calculations is a very active field where hand computation are still
necessary to find approaches suitable for automatization. These
calculations imply dealing with quite complex
renormalization and regularization problems.
However, many tools are being developed, many partial results have been
obtained
and
systematic approaches begin to emerge.
Higher loop calculations are still in their infancy,
no automatization have been yet tried, only computations with the help of
          algebraic tools has been achieved.
Complete automation are probably impossible\fnm{\dag}\fnt{\dag}{See D.
Broadhurst's paper in
these proceedings}

\subsection{Integration and event generation}
\noindent
The matrix element function is integrated over the multi-dimensional
phase space restricted by the cuts introduced by the experimental
acceptance and by the intrinsic constraints needed to tame the gauge violation
divergences.
The mapping of the integration parameters to the physics variable is called
the kinematics. This transformation must regularize or (at least)
decorrelate singularities
(infra-red divergences, mass singularity, $\gamma$-t
channel singularities, resonance formation) to
minimize the variance in order to obtain the best numerical stability.
Each singularity is mapped to an independent variable whenever
possible. Otherwise, the phase space is split to adapt to each kinematical
regions.

Multi-dimensional adaptive integration algorithms have been
developed. In the "stratified sampling" approach, the
grid spacing is adapted to the integrand magnitude and gradient
(VEGAS\cite{vegas}, BASES\cite{bases}).
Iterative integration is then performed until the requested accuracy is
reached.

New ideas based on quadtree and simplexes partitioning\cite{ilyin}
or on wavelets\cite{wavelets}
analysis are proposed to improve the performance of this major module.

The gauge invariance is checked numerically by
selecting a point in the phase space and computing the matrix
          element for the various gauge parameters and gauge schemes
(covariant, unitary, axial gauge).

Structure functions $F(x,s)$ can be taken into account to the cost of
increasing the number of
dimensions in the integration
$\sigma(s) =\int dx F(x,s)\sigma_0(x,s)$. The kinematics becomes also more
complex due to the s-dependence of the integrand.

Parton shower and hadronization are performed by independent packages.
The partonic final state is transformed into physical particles using one of
the
many hadronization scheme including:
Independent fragmentation (COJETS\cite{cojets}),
          Color string (JETSET/PYTHIA\cite{jetset},
ARIADNE{\cite{ariadne}),
          low-mass cluster (HERWIG\cite{herwig}).
Color correlation and helicity constraint (resonance formation)
are more difficult issues and should be implemented carefully.

Event generation is obtained by sampling the
distributions computed during the integration step. SPRING\cite{bases} is an
event generator which has been
developed to use the BASES integration informations.

\section{Packages}
\noindent
To date\fnm{\dag}\fnt{\dag}{Check on these proceedings for more recent updates}
only two packages, GRACE and COMPHEP, are fully automatic: from
the process definition to the cross-section final values.

The {\bf GRACE} packages have been developed by the Minami Tateya
Collaboration\cite{minami}.
GRACE-T\cite{grace-t} is a general tree level package which
adopts a numerical approach based on
           helicity amplitudes for massive particles using the CHANEL library.
The QFD and QCD model are built-in. Anomalous couplings
           and MSSM are almost ready to be released.
Integration and event generation rely on the BASES/SPRING package.
GRACE-L1 is the one-loop extension to the previous package.
The generation of n-loop and counter terms is performed, but the rest of the
package is limited to 1-loop and 4 external particles.
Algebraic computation is performed on the product of the loop and
           the tree level diagrams using {\it Reduce}. Then
a numerical integration involving scalar loop integrals is carried out.
Renormalization is performed automatically in
           4+$\epsilon$ dimensions.
$e^+e^- \to t \bar{t}$ and $\gamma \gamma \to t \bar{t}$ are some of the
processes which have been calculated with this package.
GRC++ is an interactive package based on KUIP embedded in PAW++.
It is fully interactive for up to 2$\to$4 and some 2$\to$5
processes(fig.\ref{fig:grc++}).
\begin{figure}
\vfill \begin{minipage}{.495\linewidth}
\flushleft{\fbox{\epsfig{file=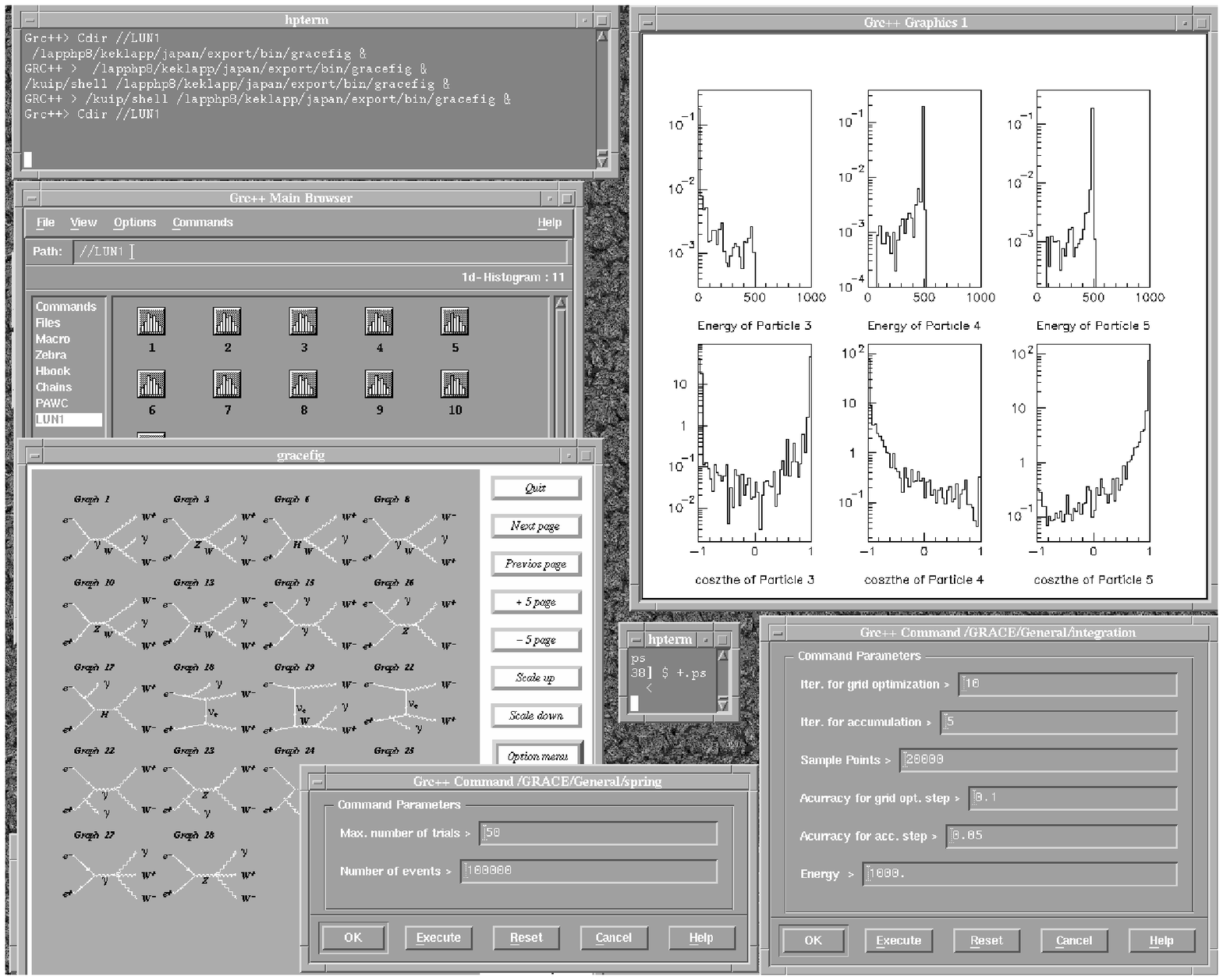,
height=6cm}}
\vspace*{0.15cm}
\fcaption{A Grc++ example} \label{fig:grc++}
}
\end{minipage} \hfill
 \begin{minipage}{.495\linewidth}
\flushright{\fbox{\epsfig{file=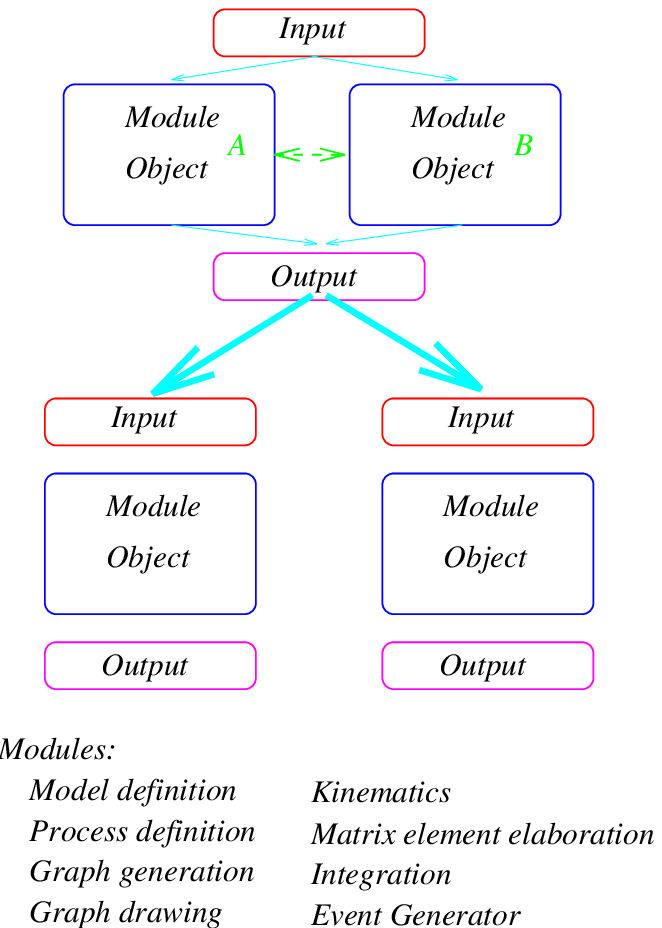,
height=6.8cm}}
\vskip 10pt
\fcaption{Standardization: an Object Oriented Approach} \label{fig:stand}
}
\end{minipage}
\vfill
\vspace*{-0.5cm}
\end{figure}

The {\bf COMPHEP} package has been developed by the INP Moscow
Group\cite{moscow}
It is a general tree level system (up to 2 $\to$ 4 process).
It is based on an analytical approach where the square of the sum of the
diagrams are computed using their own
         symbolic calculation package,{\it Reduce} or {\it Mathematica}.
The complete QFD model is treated in unitary and 'tHooft-Feynman gauge.
         Anomalous coupling and MSSM are in preparation. Furthermore
the user may define its own model.
Structure functions are implemented using PDFLIB\cite{pdflib}.
BASES and SPRING are used for integration and event
           generation. The kinematics is generated automatically for the most
current needs.
The package provides an interactive computation for the 2$\to$3 and some
2$\to$4 processes:
        from process specification to energy dependent total cross-sections
         and various angular distributions.

{\bf MADGRAPH} is an automatic diagram generation\cite{madgraph}package
 for tree level processes. It produces a matrix element in term of massless
 or massive helicity amplitudes
using the HELAS library. The QFD and QCD model are available,
MSSM will be released soon.
BRS gauge invariance check is automatically built by the package.
Diagram generation is interactive but the
         kinematics, integration and event generation are let to the user.

{\bf FEYNARTS/FEYNCALC} from
the Wurzburg Group\cite{feynart} is a {\it Mathematica} package providing
convenient tools for radiative corrections in the Standard model. It
generates and computes all tree, 1-loop graphs and some 2-loop processes. It
can handle up to 4 external particles (2$\to$2 at 1-loop and 2-loop
         photon self energy).
QFD and QCD are basically built in and MSSM is in preparation.
The 1-loop $e^+e^- \to HZ^0$ calculation\cite{denner}
has been performed with this package.

{\bf FDC},
the Feynman Diagram Computation\cite{fdc} package performs
automatic calculation of 1-loop diagrams using the Wick's theorem.
It follows an analytic approach based on {\it Reduce, RLisp} to produce the
matrix element. This package is still in development stage.

\section{Computing Aids}
\noindent
Some tools have been developed for more specific computations and are sometimes
more advanced
that some of the modules of the complete packages. However they	cannot be used
blindly to automatically produce cross-sections and event generators.

{\bf HIP\cite{hip}} is a
high level {\it Maple} or {\it Mathematica} functions library to perform
symbolic
calculation on trace of product of Dirac Matrix. It provides an easy
way to elaborate cross-section integrand and decay width functions.
It implements the Vector Equivalence technique where pair of
external fermions are replaced by an equivalent four-vector.

{\bf TRACER} from
the Garchin group\cite{tracer} is a {\it Mathematica} package to perform
symbolic manipulation and trace operations on string of
$\gamma$-algebra objects in n-dimensions using the 't Hooft-Veltman scheme.

{\bf COMPUTE\cite{compute}} from
the Raleigh Group\cite{compute} is a {\it Maple} or {\it Mathematica}
package implementing
spinor techniques for exclusive processes in perturbative QCD,
for example: Nucleon compton scattering (378 basic Feynman Diag.).

{\bf MINCER\cite{mincer}} is a  {\it Form} package dedicated to the
calculation of  massless 1, 2, 3-loop diagrams of propagator type,
for example: $Z^0$ $\to$ Hadrons ($\alpha_s^3$, NNL approximation).

{\bf PHYSICA}\cite{physica} is a {\it Mathematica} package for the
symbolic calculation of tree level processes.

\section{Standardization}
\noindent
It was decided during the workshop to set up a standardization agreement
in order to open these packages to the outside world and to fix
the structure of the event generators. Working groups have been formed and the
PISA-I
document will be available soon on http://lapphp0.in2p3.fr/aihep/aihep.html

{\bf General package structure}:
A unique definition of
the input and output of each major module should permit
the building of a unified workbench (Fig.\ref{fig:stand}) where modules could
be exchanged easily
and where the same output could serve as input to several other modules.

{\bf General structure of the event generator}:
The event generator routine needs a deeper level of standardization in order to
be interfaced to various structure functions, radiative corrections and
hadronization
packages. Moreover the general structure should permit a simple
introduction in the large detector simulation packages.

\section{Conclusions}
\noindent
Complete tree level automatic computation systems like GRACE, COMPHEP and
MADGRAPH, complementary in their approaches, are
available today.
Improvements in the code efficiency, in the user interface and in the
kinematics libraries are been pursued.
Based on these packages, cross-sections and event generation databases are
being prepared\cite{minami} \cite{moscow} for all tree
level processes($2 \to 2,3,4,5$).
Complete 1-loop QFD computations are close to
be released. Although numerical stability problems and limitation to
4 external particles may reduce its practical use,
it clearly demonstrate the feasibility of this programme and,
judging from the work being dedicated to these issues, it will keep improving
and getting more general in the coming years.
QCD packages are now being actively developed. Next to leading order
computations
are investigated in the framework of supersymmetric and string based
approaches.
Complete automatic 2-loop correction packages are still in the science fiction
section
as they are one
order of magnitude higher in technical complexity and
in the need for computer performances. However a lot of activity is devoted to
these themes and progress will definitely come.
Complete automatic n-loop calculation seems out of reach with today technology.
The future of automatic Feynman diagrams automatic computation depends
on a widely accepted standardization scheme. Working groups
have been created during this workshop to produce the
first first Standardization Agreement document (PISA-I) very soon.

\nonumsection{Acknowledgments}
\noindent
I am happy to thank all my colleagues from the LAPP theory group,
from the Minami Tateya collaboration and from the Moscow State University (INP)
for their constant help and support during the preparation of this talk.

\end{document}